\documentclass[a4paper,11pt]{article}

\usepackage{makeidx,url}
\usepackage{mathptmx}       
\usepackage{helvet}         
\usepackage{courier}        
\usepackage{type1cm}        
\usepackage{makeidx}         
\usepackage{graphicx}        
\usepackage{multicol}        
\usepackage[bottom]{footmisc}
\usepackage{amsmath,amssymb}
\usepackage[authoryear]{natbib}
\usepackage{natbibspacing}

\def\sect{{Section}}
\def\msun{{M$_\odot$}}
\def\cc{{cm$^{-3}$}}
\def\kms{{km\,s$^{-1}$}}
\def\arcmin{{$^\prime$}}
\def\arcsec{{$^{\prime\prime}$}}
\def\deg{{$^{\circ}$}}

\def\aap{A\&A}

\def\apj{ApJ}
\def\aj{AJ}
\def\apjl{ApJ}
\def\apjs{ApJS}
\def\pasj{PASJ}

\def\ssr{Space Science Rev.}
\def\mnras{MNRAS}
\def\araa{Ann. Rev. A\&A}

\def\nat{{Nature}}

\begin{document}

\title{Supernova 1604, Kepler's supernova, and its remnant}

\author{Jacco Vink\\
University of Amsterdam,
  Science Park 904, 1098 XH Amsterdam\\
  j.vink@uva.nl
}
\date{}
\maketitle

\section*{Abstract}
{\footnotesize
Supernova 1604 is the last  Galactic supernova for which historical records exist.
Johannes Kepler's name is attached to it, as he published
a detailed account of the observations made  by himself and European colleagues.
Supernova 1604 was very likely a Type Ia supernova, which exploded
 350~pc to 750~pc above the Galactic plane.
 Its supernova remnant, known as Kepler's supernova remnant, 
 shows clear evidence for interaction with
 nitrogen-rich material in the north/northwest part of the remnant,
 which, given the height above the Galactic plane, must find its
 origin in mass loss from the supernova progenitor system.
 The combination of a Type Ia supernova and the presence
 of circumstellar material makes Kepler's supernova remnant
 a unique object to study the origin of Type Ia supernovae.
 The evidence suggests that
  the progenitor binary system  of supernova 1604
 consisted of a carbon-oxygen white dwarf and an evolved
 companion star, which most likely was  in the (post) asymptotic giant branch of its evolution.
 A problem with this scenario is that the companion star must have survived
 the explosion, but no trace of its existence has yet been found, despite a deep search.
 }
 \tableofcontents

 \section{Introduction}
\label{sec:intro}

Supernova (SN) 1604 \index{SN\,1604 (SNR G4.5+6.8)} is the last of the historical supernovae. We know of at least
two other supernovae with later explosion dates, those associated
with the supernova remnants Cassiopeia A     \index{Cassiopeia A (Cas A, G111.7-2.1)}  \index{SNR G111.7-2.1 (Cas A)}
and G1.9+0.3 \citep{green08}, \index{SNR G1.9+0.3}
but no known historical records of the associated supernova events exist.

The discovery of SN\,1604 occurred on October 9, 1604 (Gregorian calendar),
early in the evening when all interested in astronomy gazed the sky
to watch the conjunction of  Mars, Jupiter and Saturn.
This special circumstance probably helped the early
discovery of the supernova at around 
20 days before
maximum brightness \citep{stephenson02}.
The occurrence of a new star in conjunction with three bright planets 
lead to fierce debates among contemporary astronomers,
most of whom still held partial medieval world views, with many elements
of astrology. Of all these astronomers, it is in particular
Johannes Kepler's \index{Kepler, Johannes} name that is now attached to this supernova, 
because  he wrote a book ``De Stella Nova in Pede Serpentarii'' ("On the new star in the foot of the Snake [Ophichius]")
about the new star,
in which he
published his own observations of the supernova and those by various European  colleagues,
and discusses the importance of the new
star, including a possible link to the \index{star of Bethlehem} star of Bethlehem.
Kepler only started observing SN 1604 relatively late, October 17, due to unfortunate  weather 
circumstances in Prague, his city of residence.
Kepler and his contemporaries attached a great deal of metaphysical significance to the appearance of
the new star: {\em ``The star's significance is a difficult matter to establish and we can be sure of only one thing: 
that either the star signifies nothing at all for Mankind or it signifies something of such exalted importance that it is beyond the grasp and understanding of any man''}.
Kepler himself, in fact, was a bit more cautious about the metaphysical implications of the event than some of his contemporaries \citep{granada05}.

Currently, the  remnant of SN 1604, labeled G4.5+6.8, \index{SNR G4.5+6.8} but
usually called \index{Kepler's supernova remnant (SNR G4.5+6.8)} Kepler's supernova remnant (SNR), or even "Kepler" for short,
is one of the best studied SNRs.  It appears that Kepler's SNR has indeed a wide ranging significance,
but not for metaphysical reasons, but  for understanding the nature of  \index{Type Ia supernova} Type Ia supernovae.
SN 1604 was almost certainly a  Type Ia supernova, but, peculiarly, the SNR itself is interacting
with a nitrogen-rich \index{CSM (circumstellar medium)} \index{circumstellar medium} circumstellar medium (CSM) at a distance of 2-3 pc from the explosion centre, 
suggesting that the progenitor system had significant \index{stellar mass loss} \index{stellar wind} mass loss. 

As described in more detail in Sections 3 and 6, two basic types of supernova Type Ia progenitor models exists  \index{single degenerate Type Ia model} \index{double degenerate Type Ia model} 
\citep{maoz14}: 1) the single degenerate model,
according to  which a CO white dwarf \index{white dwarf} accretes  from an (evolved) stellar companion until the pressure
in the core leads to the thermonuclear explosion \index{thermonuclear explosion} of the white dwarf, at the moment that the progenitor
has  a mass approaching the \index{Chandrasekar mass} Chandrasekar mass (1.38 \msun), and 2) the 
double degenerate model, according to which Type Ia explosions are triggered by the merging of two white dwarfs.
The time scale from birth of the binary system to the merging of the two white dwarfs is generally expected to  be a few $10^9$~yr \citep{maoz14}, 
since the two stars are brought closer to each other due  gravitational radiation.
The fact that Kepler's SNR is interacting with CSM is, therefore, more consistent
with the single degenerate scenario. However, as will be discussed, this scenario is also not fully consistent
with the properties of Kepler's SNR, in particular with the lack of a surviving donor star.

\begin{figure}
\includegraphics[trim=0 300 0 0,clip=true,width=\textwidth]{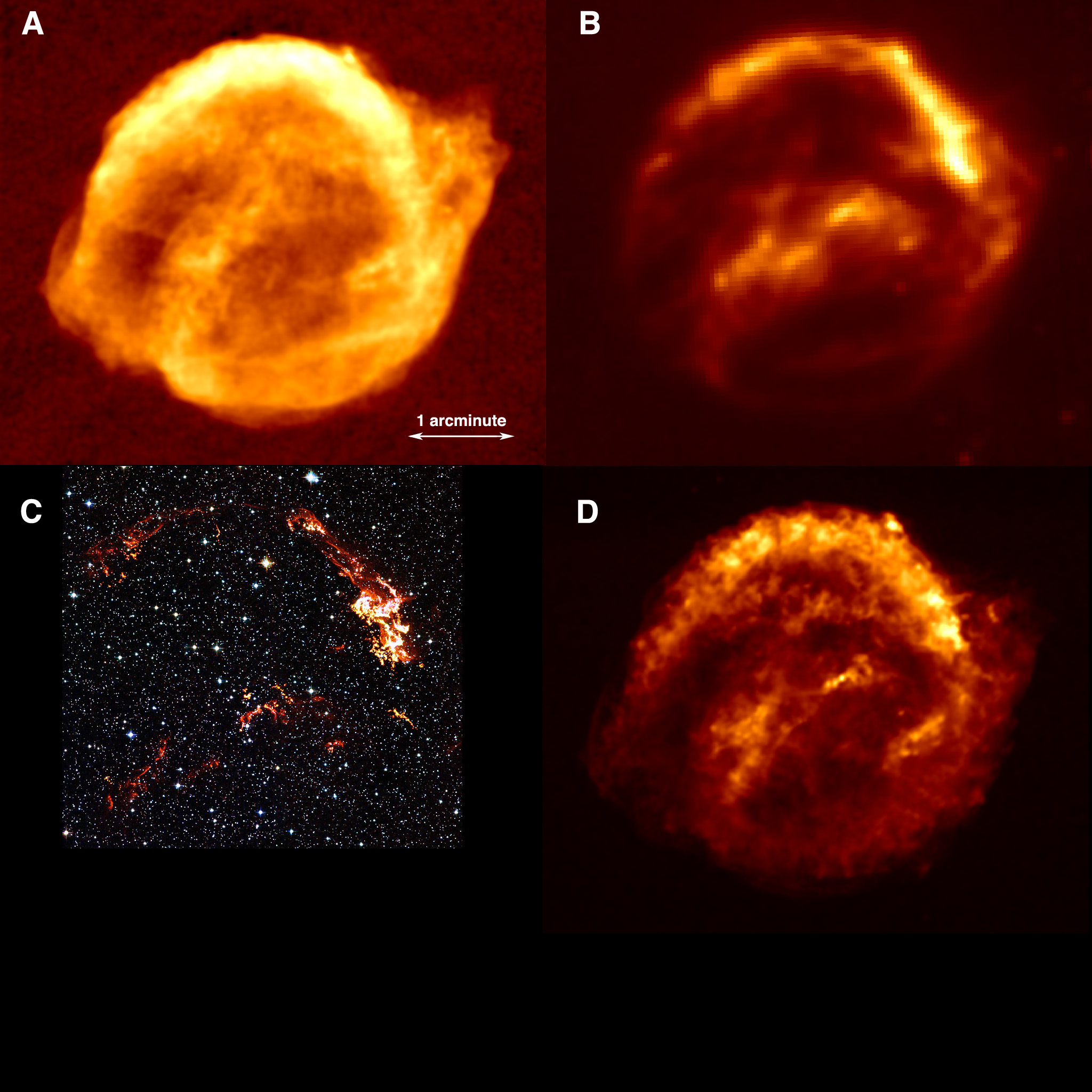}
\caption{\label{fig:4images}
Kepler's SNR at different wavelengths: a) Radio VLA (4.85 GHz) map \citep{delaney03}; b) 24$\mu$m dust emission as
observed by the Spitzer MIPS instrument \citep{blair07}; c) optical image (H$\alpha$, [NII] and  [OIII]) obtained
by the Hubble Space Telescope/ACS instrument (credit: ESA/NASA, The Hubble Heritage Team (STScI/AURA));
d) Chandra X-ray image in the iron-rich 0.7-1.0 keV band \citep{reynolds07}.
Intensity scaling of Fig. a, b, and d are square-root scalings, bringing out fainter details.
}
\end{figure}

\section{The supernova remnant, its distance and its multiwavelength properties}
\label{sec:properties}

\subsection{Position, distance estimates and SN 1604 as a runaway system}

Kepler's SNR was discovered by Walter Baade  \index{Walter Baade}  as "a small patch of nebulosity close to the expected
place" \citep{baade43}. \citet{minkowski59} revealed that the spectrum of the optical nebula displayed strong [NII]
emission as compared to H$\alpha$ emission. 
The H$\alpha$/[NII] is concentrated toward the north/northwestern region of the SNR and a bar-shape
region across the center, roughly oriented in the NW-SE direction (Fig.~\ref{fig:4images}).
The overall extent of the SNR is more clear from radio and X-ray images of the SNR, revealing
a roughly spherical shell with two protrusions ("Ears") in the NW and SE \citep[e.g.][]{delaney03,reynolds07}.
The SNR's center is located at $\alpha_\mathrm{J2000}=17{\rm h}30{\rm m}41{\rm s} ,\ 
\delta_\mathrm{J2000}=-21^\circ 29^\prime 32^{\prime\prime}$, in Galactic coordinates
$l=4.52^\circ, b=6.82^\circ$. The angular radius of the SNR is approximately 1.8\arcmin\ (not taking
into account the "Ears").

The distance to the SNR is uncertain, with estimates varying from $d=3.2$~kpc \citep{danziger80} to  
$d=12$~kpc \citep[][]{vandenbergh73}. 
The long distance was based on the maximum visual magnitude of the supernova event of
$V_\mathrm{max}\approx -2.5$, as estimated by \citet{baade43} from the historical records.
After correcting for absorption \citep[$A_\mathrm{V}=2.17-3.47$,][]{vandenbergh77,danziger80}, the peak
magnitude can then be compared to the  absolute magnitude of Type Ia supernovae of $M_\mathrm{V}=-19.0 \pm 0.5$.
A Hubble parameter was adopted of $H_0=100$~km/s/Mpc.

More recent distance estimates, not based on the historical peak brightness \index{peak brightness} of the
supernova,  vary widely, ranging from $d=3.9\pm 1.4$~kpc \citep{sankrit05}
based on combining proper motion and Doppler broadening of H$\alpha$ emission, 
 up to, or even beyond, 6-7~kpc based on the angular size of
the SNR, combined with the energetics of the supernova \citep[][see \sect~\ref{sec:dynamics} for details]{aharonian08,chiotellis12,patnaude12}.
Radio absorption measurements
toward the SNR are consistent with a distance in the range 4.8-6.4~kpc \citep{reynoso99}.

The Galactic latitude of Kepler's SNR ($b=6.8$\deg) implies a large height above the Galactic plane
of $z=594 d_5$~pc (with $d_5$ the distance in units of 5~kpc). The large separation from  active star
forming regions suggests that SN 1604 was Type Ia supernova, in agreement with the reconstructed
supernova light-curve, as already pointed out by \citet{baade43}. However, the large amounts of
nitrogen present in the optical nebula can only have come from material processed by the CNO nucleosynthesis
cycle, suggesting a  stellar wind origin for the optical nebula. This lead \citet{bandiera87} to suggest  that
the progenitor was a massive star, which escaped the Galactic plane with a high velocity. The high
proper motion of the progenitor is in agreement with observations of the kinematics of the optical nebula
(\sect~\ref{sec:dynamics}). Given the historical light curve, 
a Type II origin of SN 1604 was excluded, but a Type Ib \index{Type Ib supernova} supernova
was consistent with both the light curve and the idea of a massive runaway star as the progenitor of the supernova \citet{bandiera87}.

\begin{figure}
\centerline{
\includegraphics[width=0.85\textwidth]{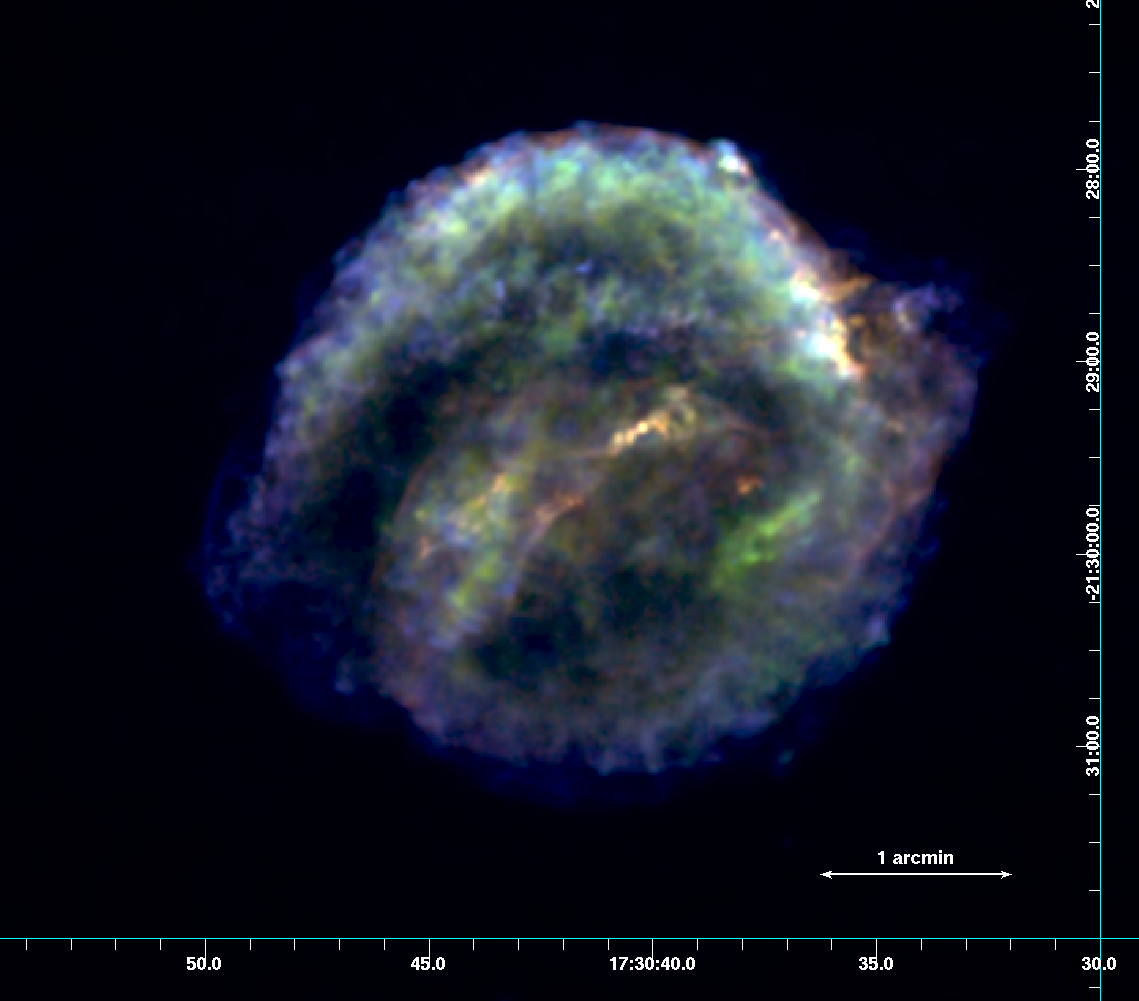}
}
\caption{
Chandra X-ray image, with the three image channels  (red, green, blue) 
corresponding
to oxygen emission (0.5-0.7 keV), Fe-L emission \index{Fe-L complex} (0.7-1 keV) and Si-K 
emission (1.7-1.9 keV), respectively. It can be seen that the Si-K emission (blue)
peaks more outward than the the Fe-L shell emission (green). Peaks in the oxygen emission
(red/yellowish) occur at the outskirts in the northeast and along the bar, suggesting
that the emission comes from shock-heated, circumstellar material.
\label{fig:Xray}}
\end{figure}

\begin{figure}
\centerline{\includegraphics[width=0.8\textwidth]{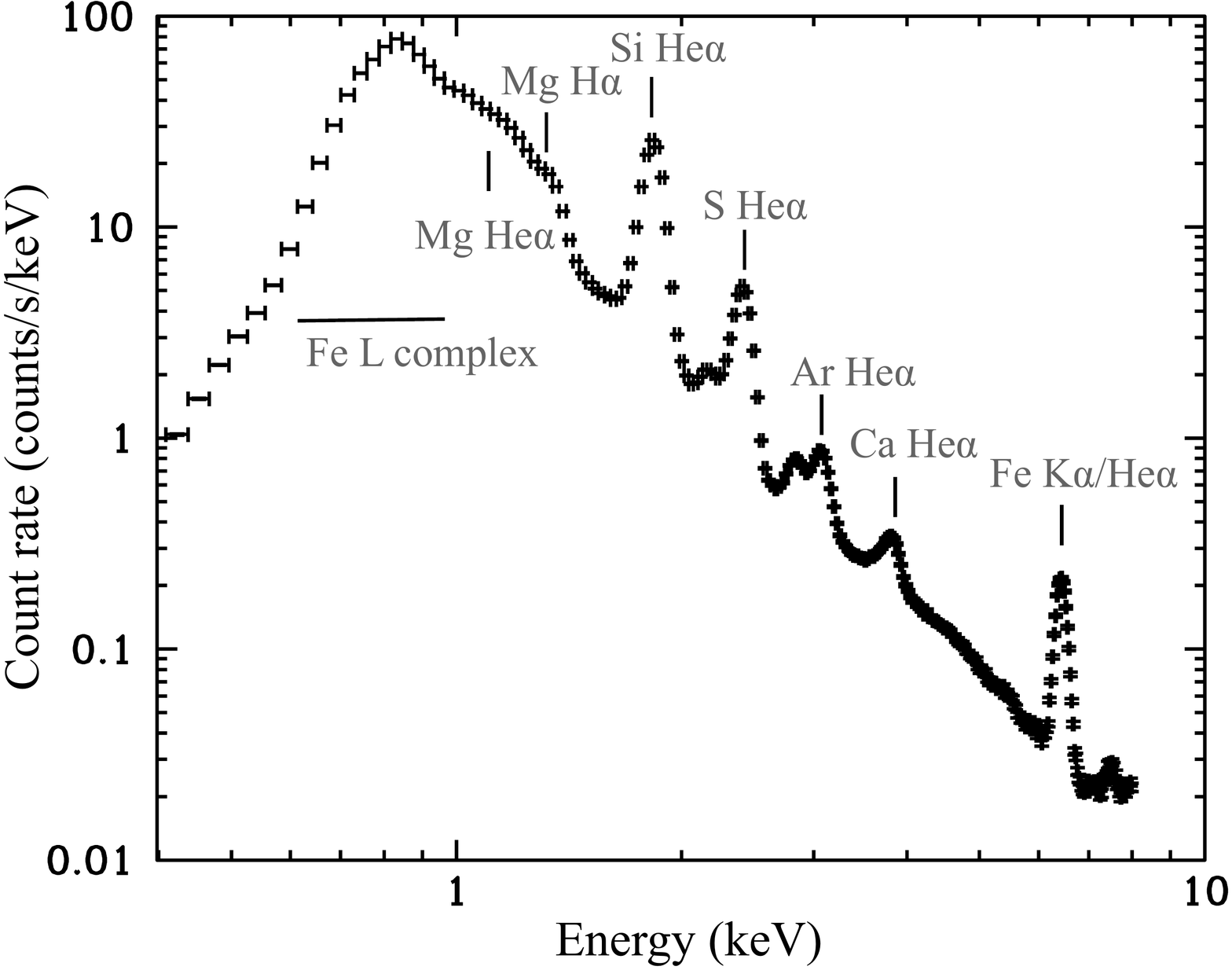}}
\caption{
The Chandra ACIS-S X-ray spectrum of Kepler's SNR, observed in 2006.
The "bump" between 0.7- 1.2 keV consists mainly of Fe-L emission \index{Fe-L complex}, consisting
of complex of lines from Fe XVII to Fe XXIV. The prominence of this Fe-L complex
in the spectrum of Kepler's SNR is indicative of a Type Ia origin.\label{fig:xspec}
}
\end{figure}

\begin{figure}
\centerline{
\includegraphics[width=0.85\textwidth]{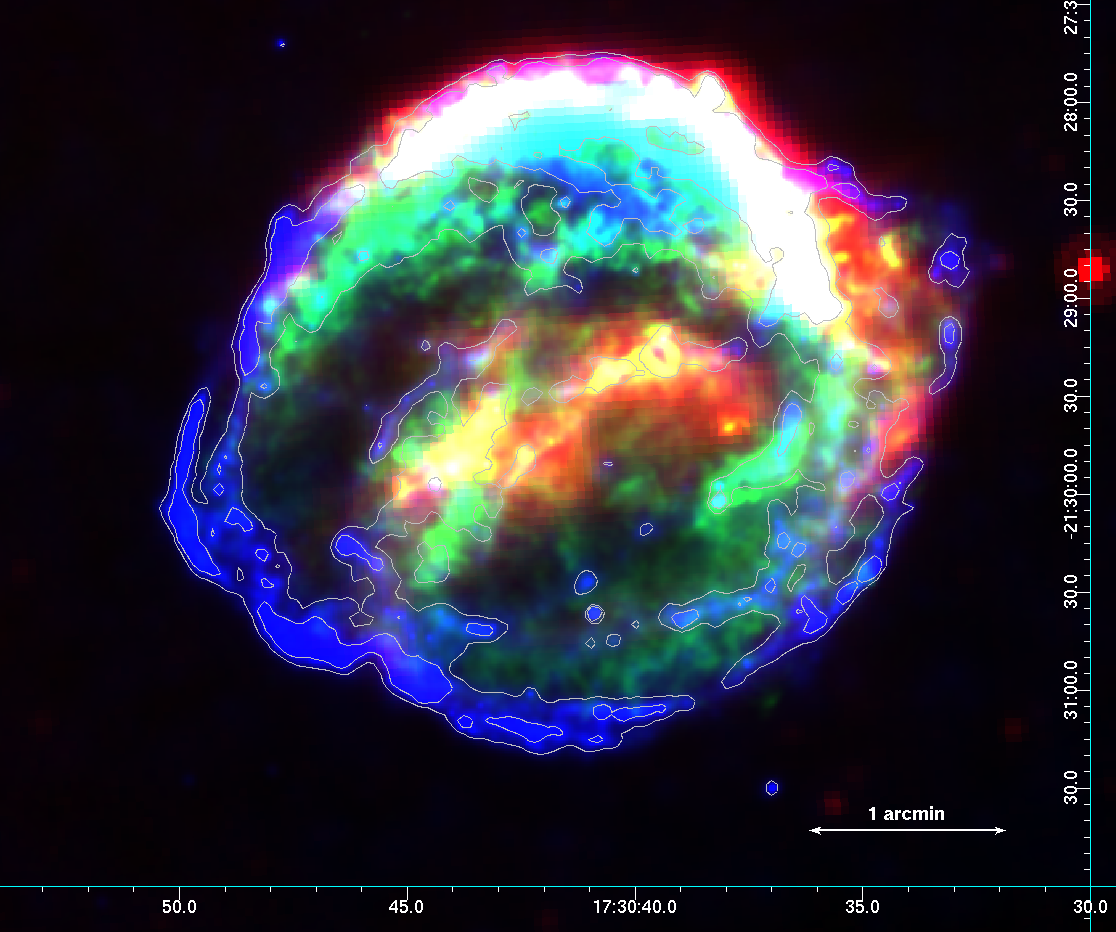}
}
\caption{
Multiwavelength image of Kepler's SNR. The red channel represents the 24$~\mu$m dust emission as
observed by the Spitzer MIPS instrument \citep[][]{blair07}. The green channel is the Si-K band (1.7-2.0 keV), representing
the bulk of the thermal X-ray emission, associated with Si-rich ejecta. The Si-K emission comes  on average from a smaller radius than the dust emission,
indicating that the dust originates from the swept up circumstellar medium, rather than the ejecta.
The blue channel (enhanced with contours) shows the X-ray continuum emission in
the 4-6 keV band. In the SE the emission is dominated by synchrotron emission from close to the shock front. \index{X-ray synchrotron radiation}
In the NW it is probably a mixture of synchrotron emission and thermal bremsstrahlung.
The X-ray bands are taken from the 2006 Chandra observations of Kepler's SNR \citep{reynolds07}.
\label{fig:morphology}}
\end{figure}

\index{X-ray imaging spectroscopy}
\subsection{X-ray imaging spectroscopy and SN 1604 as a Type Ia supernova }
\label{sec:xrays}
The idea of a Type Ib supernova origin for SN 1604 was discredited with the advent of X-ray imaging
spectroscopy with CCDs. These type of detectors were first used by
\index{ASCA X-ray satellite} ASCA, and it provides now still the most often used instrumentation for X-ray observatories
such as \index{Chandra X-ray observatory} \index{XMM-Newton X-ray space observatory} 
\index{Suzaku X-ray satellite} 
Chandra, XMM-Newton, and Suzaku. 
The X-ray spectrum of Kepler's SNR (Fig.~\ref{fig:xspec}) shows that the
dominant line emission is caused by  \index{Fe-L complex} Fe-L transitions \citep{kinugasa99,cassam04,reynolds07}, which is very indicative of
Type Ia SNRs, since they produce 0.3-1.3 \msun\ of Fe and Fe-group elements. 
In addition line emission from intermediate mass elements \index{intermediate mass elements}
like silicon, sulphur, argon and calcium, are prominent.
Unlike Kepler's SNR, young core collapse SNRs on the other hand, show dominant emission from oxygen, neon and magnesium.
Recently, \citet{katsuda15} confirmed the high nitrogen \index{nitrogen} abundance in Kepler's SNR using high resolution X-ray spectroscopy with the XMM-Newton reflective
grating spectrometer.

The spatial distribution of the  X-ray line emission indicates that the ejecta are layered according to element mass, with the intermediate mass elements 
located, on average, at a larger radius than iron \citep[][illustrated in Fig.~\ref{fig:Xray}]{cassam04}. This distribution is similar to that of other Type Ia SNRs like Tycho's SNR \citep[SN\,1572,][]{hwang97} \index{SNR G120.1+1.4} \index{Tycho's supernova remnant (SNR G120.1+1.4)} 
\index{SN\,1572 (SNR G120.1+1.4)}
and
SNR  0519-69.0 \citep{kosenko10}, but Fe-L emission is more prominent in Kepler's than in Tycho's SNR, whereas SNR 0519-69.0 and
SN\,1006  display more \index{SNR\,0519-69.0} \index{SN\,1006 (SNR G327.6+14.6)} \index{SNR G327.6+14.6} 
oxygen line emission from the outer layers.
The relatively high iron abundance in Kepler's SNR,
as compared to other Type Ia SNRs, suggests that SN 1604 was an iron/nickel-rich supernova, similar to the peculiar Type Ia supernova SN1991T \citep[][but see Section~\ref{sec:lightcurve}]{patnaude12}.

Recently it has become possible to detect line emission from the uneven elements manganese (Mn) and chromium (Cr).
The Mn yields for supernovae are a result of the presence of neutron-rich nuclei in the progenitors, which ultimately is linked to the nitrogen abundance of
the progenitor at the main sequence. Hence, it can be used to infer the metallicity of the progenitor. \citet{park13} used the Mn/Cr ratio to show
that SN 1604 must originate from a star with super-solar metallicity, ruling out a halo star progenitor.

The continuum X-ray emission of Kepler's SNR, best distinguished in the 4-6 keV band (blue in Fig.~\ref{fig:morphology}), shows narrow filaments tracing the outer region of
the SNR. \index{X-ray synchrotron filaments} Most of this emission, especially the narrow filaments in the SE,
 are thought to be caused by X-ray synchrotron emission from
10-100 TeV shock accelerated electrons \citep[][for a review]{helder12b}. The widths of the filaments can be used to estimate the post-shock 
magnetic fields, indicating for Kepler's SNR a magnetic field strength of $B\approx 120$~$\mu$G. Since the 10-100~TeV lose their energy very fast in such  a magnetic field ($\sim 10-20$~yr), the X-ray synchrotron radiation is an excellent tracer of  the shock front. Some of the X-ray synchrotron filaments
are situated along the edges of the central bar (Fig.~\ref{fig:morphology}), suggesting that we observe the shock front  under a
(nearly) edge-on viewing angle. Given the position of the bar in the center of the SNR, this  viewing angle  indicates  
a peculiar, perhaps halter-like, morphology of the SNR in three dimensions \citep{burkey13}.

\subsection{The circumstellar medium as studied in the optical and infrared}
\label{sec:opt_ir}
The shocked interstellar medium in Kepler's SNR  is best studied in the optical and infrared. The optical emission, dominated by
hydrogen line emission, comes from northwestern part of the SNR and from the central bar. \index{radiative shocks} \index{non-radiative shocks}
The optical emission can be divided in emission caused by either radiative or non-radiative shocks \citep{dennefeld82,fesen87,blair91,sankrit08}. 
Radiative shock emission is characterised by a mix of hydrogen line emission (Balmer and Lyman lines) and
forbidden line emission from NII, OIII, and SII. The H$\alpha$ line emission is in those cases  double peaked, since NII has among
others line emission at $\lambda=654.9$~nm, close to H$\alpha$ ($\lambda=656.4$~nm).
Forbidden line emission comes from regions heated by slow shocks ($\lesssim 200$~\kms), 
which heat plasma to temperatures below $\sim 10^6$~K. For this temperature regime radiative losses by optical/UV line emission causes rapid cooling. 
For Kepler's SNR the overall shock speed is several thousand \kms. 
So the radiative emission must come from dense knots \citep[$n \gtrsim 10^3$~\cc][]{leibowitz83}, in which locally the shock wave
has decelerated considerably. The [NII]/hydrogen line ratios measured for radiative shocks in Kepler's SNR 
suggests an overabundance in nitrogen of a factor 2-3.5  \citep{leibowitz83,blair91}.

The non-radiative emission is caused by neutral hydrogen atoms entering the shock, after which they may not be immediately ionised, but first
get into an excited state, or undergo will charge exchange  with shock-heated protons behind the shock front. H$\alpha$ line spectra of 
non-radiative shocks typically consist of  narrow-line emission caused by direct excitation of the neutrals, and broad-line emission caused by
hot protons that just picked up an electron through charge exchange.
The narrow line width and line center can be used to measure the plasma speed
and temperature of the un-shocked gas, whereas the broad-line width can be used to measure the temperature immediately ($\lesssim 10^{15}/n_\mathrm{p}$~cm)
behind the shock (see \sect~\ref{sec:dynamics}). The H$\alpha$ flux of the non-radiative filaments can be used to infer densities as well,
which appears to be of the order of $\sim 10$~\cc \citep{blair91}. This is surprisingly large, given the height above the Galactic plane
of $594 d_5$~pc. 
The optical emission from Kepler's SNR, therefore indicates that the SNR is interacting
with  dense, CNO processed material, which, given the isolated nature of the SNR high above the Galactic plane, 
can only be understood by assuming that the circumstellar material originates from 
the progenitor system of SN\,1604.

In the infrared Kepler's SNR has been observed in the wavelength range from 12-100$\mu$m by all major
infrared observatories \index{Infrared Astronomical Satellite (IRAS)} \index{Spitzer Space Telescope} \index{Herschel Space observatory} \index{Infrared Space Observatory (ISO)}
\citep[IRAS, ISO, Spitzer, Herschel, see][]{braun87,douvion01,blair07,gomez12,williams12}.
Additional emission for wavelength larger than 100~$\mu$m, even up to 850~$\mu$m may be present \citep{morgan03,gomez12}, but
at long wavelengths background subtraction is difficult, and even at 100~$\mu$m there is already a factor three discrepancy
between IRAS and Herschel based flux measurements \citep{gomez09}.

\index{dust emission}
Infrared emission from SNRs is caused by warm dust, heated by collisions with hot electrons and ions. The dust temperature
 is established by the equilibrium between collisional heating and thermal emission \citep[][]{draine81,dwek87}. 
 Both heating
and emission depend on the surface area of the dust grains, and for large dust grains the temperature is independent of the size
of the dust particles. However, since the wavelength of the radiation and the dust particle size can be comparable, the dust emission
cannot be adequately described by Stefan-Boltzmann's law. Moreover, the emission spectrum also depends on the composition of the dust.

Infrared photometry of Kepler's SNR indicates dust temperatures of 80--120~K \citep{braun87,douvion01,blair07,gomez12,williams12},
whereas more detailed spectra obtained with the ISO-SWS and Spitzer-IRS spectrometers  \citep[respectively][]{douvion01,williams12} 
indicate that the dust particles themselves are  silicates (silicon oxides). Interestingly, the modelling of the spectra gives rise
to different interpretations. \citet{douvion01} conclude that the dust comes from regions with densities of $n_\mathrm{e}=500-7500$~cm$^{-3}$
with plasma temperatures of $(0.4-6)\times 10^5$~K. On the other hand, \citet{williams12} estimates the  densities to be  $n_\mathrm{e}\approx 5$~cm$^{-3}$ in the South
to $n_\mathrm{e}\approx 50$~cm$^{-3}$ in the North, but assumes that the plasma temperatures are consistent with the X-ray measured
temperatures ($T_\mathrm{e}\approx 10^7$~K, and  ion temperatures a factor ten higher). Both conditions may arise
in Kepler's SNRs as the radiative shocks are going through very dense regions, whereas the non-radiative shocks are responsible
for the X-ray emitting plasma. Indeed, several of the regions analysed by \citet{williams12} contain non-radiative shock emission. The infrared spectrum
of the region in which radiative shock emission is dominant is characterised by bright forbidden line emission in the infrared (among others  [Si II], [Fe II]). 
It is  important to note  that the location of the dust emission indicates that
all the dust is associated with shocked CSM,
and not with supernova ejecta. This appears to be a characteristic of Type Ia supernovae, which do not seem to produce dust in their ejecta \citep{williams12}.
This is in contrast to a core collapse SNR like Cas A \citep{lagage96} or SN 1987A, where dust clearly is formed from
supernova ejecta. \index{Cassiopeia A (Cas A, G111.7-2.1)}  \index{SNR G111.7-2.1 (Cas A)}
\index{SN 1987A}
\index{dust emission}
\index{dust production}

\section{The dynamics of Kepler's SNR}
\label{sec:dynamics}

\subsection{Velocity measurements}
The earliest measurements of  velocities in Kepler SNR's were  radial velocity measurements of the bright optical
knots, indicating line emission that is blue shifted by about 230 \kms \citep{minkowski59}. Optical proper motion studies
\citep{vandenbergh77,bandiera91}
also indicated velocity offsets,  indicating that the bright optical knots have a bulk velocity of 212 $d_5$\kms\ in northwestern direction,
away from the Galactic plane,
whereas the expansion velocity, based on spherical expansion directed away from the centre of the SNR, is only $(74\pm 29)d_5$~\kms
\citep{bandiera91}. The slow expansion velocity, corresponding to an expansion time scale of 32000~yr, indicates that the velocity
of the optical knots reflects mostly the velocity of these knots prior to interaction with the SNR blast wave. 

The more diffuse non-radiative H$\alpha$ emission \citep{fesen87} provides information on the local shock wave velocities through
the width of the broad H$\alpha$- component; and the narrow line component
provides information on the velocity of the unshocked, partial neutral gas,
 immediately ahead of the shock.
The latter informs us that the unshocked gas is blue shifted with about $180$~\kms \citep{blair91,sollerman03}.

The width of the broad line H$\alpha$ component is caused by thermal Doppler broadening \index{Doppler broadening} of shock-heated protons.
The reported H$\alpha$ line FWHM of $\sim 1800$~\kms\ translates into  shock velocities of $1500-2000$~\kms \citep{blair91}.
Part of the uncertainty is caused by the uncertainty in electron-ion temperature ratio. \cite{sankrit05} combined this shock velocity measurement
with a measure proper motion of diffuse filament in the NW of $1.45\pm 0.03$ \arcsec/16.33 yr to infer a distance to Kepler's SNR
of 3.0-5.0~kpc. However, note that some of the Doppler broadening widths measured by  \cite{blair91} indicate 
shock velocities larger than 2000~\kms. Notably, the filament labeled "NW diffuse emission" has a width of $3409 \pm 454$~\kms, which corresponds
to a shock velocity range of 2300-4000~\kms \citep{vanadelsberg08}, or even larger if cosmic-ray acceleration \index{cosmic rays} absorbs part of the shock
energy-flux into the shock, which results in lower than expected post-shock temperatures \citep[e.g.][]{vink10a}.

\begin{figure}
\centerline{\includegraphics[trim=0 -30 0 0 0,width=0.4\textwidth]{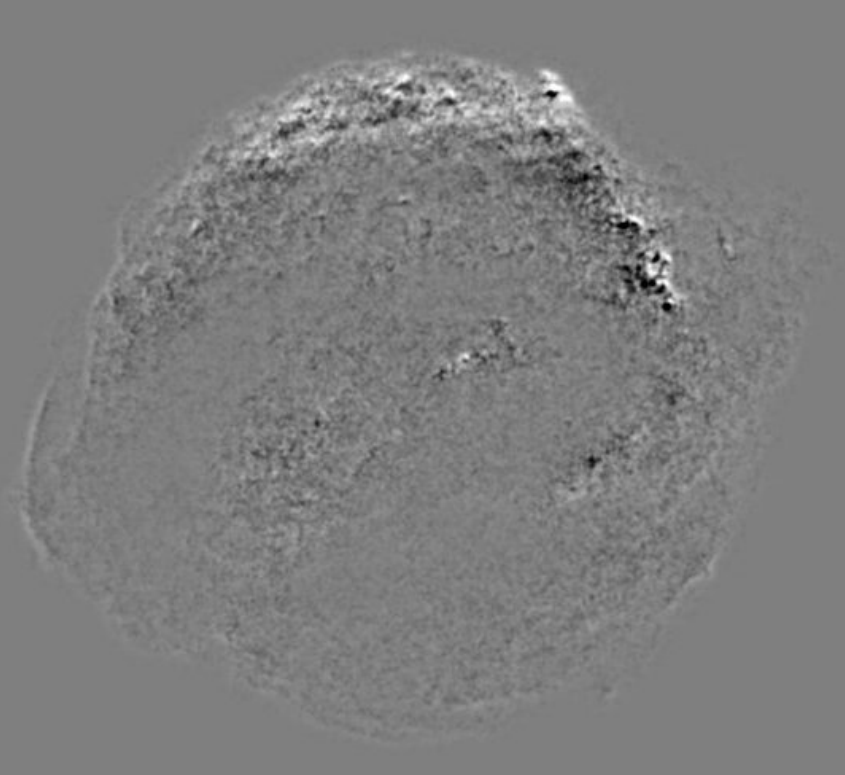}
{\includegraphics[width=0.6\textwidth]{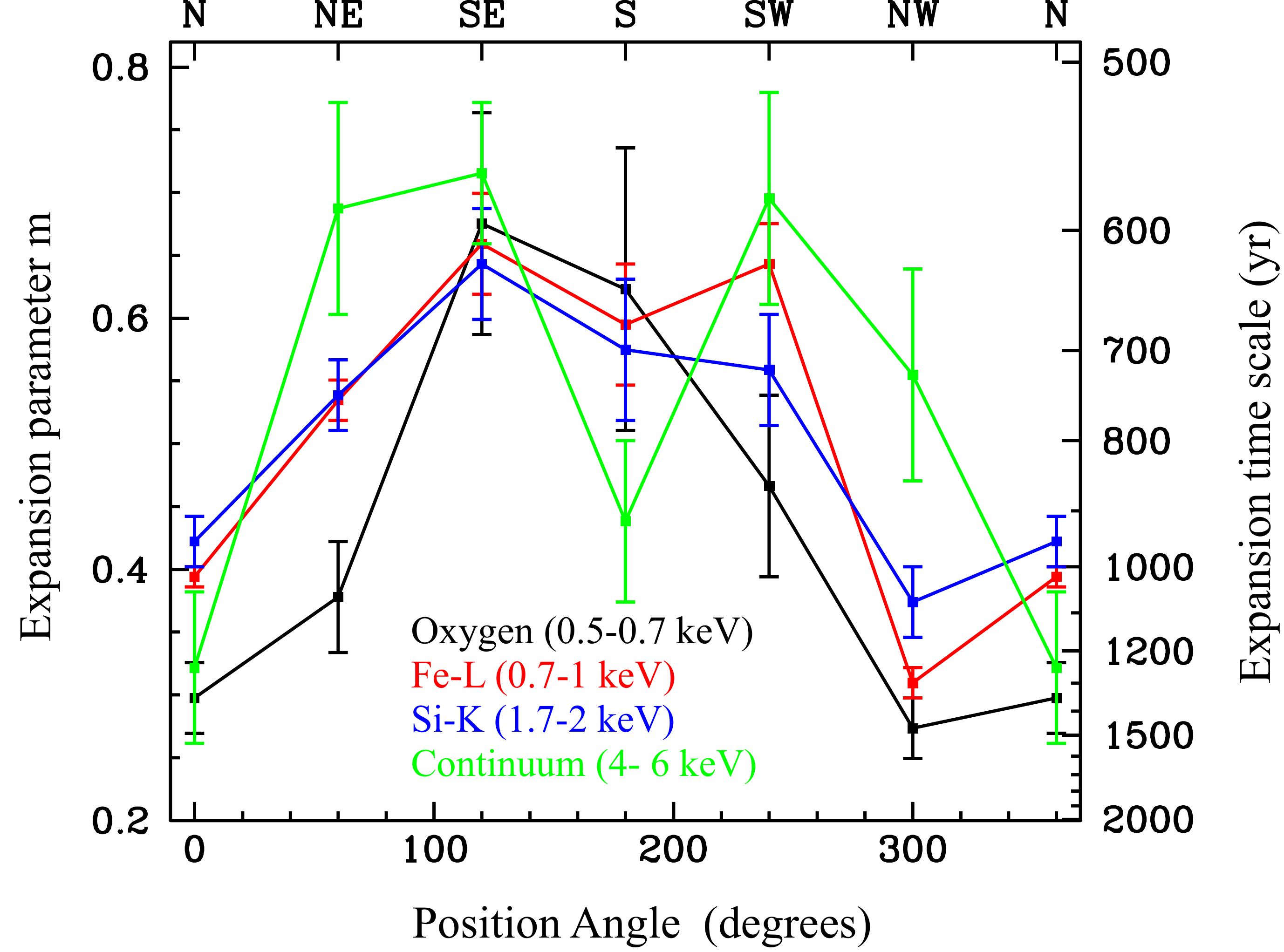}}
}
\caption{
Left: Difference image between images in the 1-1.5 keV band in 2000 and 2006, as observed by Chandra.
The shadow-like features are caused  by proper motions.
Right: X-ray expansion measurements based on these two Chandra images in various energy bands \citep[adapted from figure in][]{vink08b}.
\label{fig:expansion}
}
\end{figure}

The optical velocity measurements pertain to the densest regions of Kepler's SNR. The overall
expansion of Kepler's SNR has been measured in the radio \citep[][with VLA]{dickel88} and X-rays \citep[][see Fig.~\ref{fig:expansion}]{vink08b,katsuda08} using proper
motion measurements. These measurements show that the northwestern part is expanding at a slower rate than the southwestern part.
A useful way to characterise the expansion rate is in terms of the expansion parameter 
\begin{equation}
m \equiv \frac{R/t}{V},
\end{equation}
with $R$ the radius, $t$ the age of the SNR and $V$ the velocity. For proper motions both $R$ and $V$ depend in a similar way on the distance,
so one can replace $R$ and $V$ with angular radius and proper motion. 
One can also calculate the \index{expansion age} expansion age $\tau_\mathrm{exp}\equiv R/V$, in which case $m=t/\tau_\mathrm{exp}$, \index{expansion parameter}
with $t$ the true age
of the SNR. For SNRs in the Sedov-Taylor phase of the evolution we
expect $R \propto t^{2/5}$ and hence $m=2/5$, whereas in the ejecta dominated phase one expects $m=0.7-0.9$ \citep{chevalier82}.
It turns out that for Kepler's SNR in the Northwest $m=0.3-0.35$, whereas in the Southwest the expansion is faster, with $m=0.6-0.7$.
The expansion parameter in the northwest is lower than $m=2/5$, which indicates that the shock  must have encountered a density enhancement.
\citet{vink08b} estimates that the  total excess mass toward the northwestern region is about 1~\msun.

The proper measured proper motions translate in shock velocities of
\begin{equation*}
V_\mathrm{sh}= m 9920(\theta/2.79) d_5~\mathrm{km\ s^{-1}}
\end{equation*} 
for an age of 400~yr, with $\theta=2.79$\arcmin\ the angular radius of Kepler's SNR.
Given the measured values of $m$ the shock velocities around Kepler's SNR are, therefore, in the range of 2900-7000 $d_5$~\kms.

\subsection{Hydrodynamical simulations of Kepler's SNR}

As discussed above the density enhancements in the north-northwestern regions, as well as the evidence for high density,
nitrogen-rich material suggests that the progenitor system of SN\,1604 suffered significant mass loss. 
\citet{borkowski94} simulated this situation in the context of a Type Ib origin of SN\,1604 and \citet{chiotellis12} adapted it
for the case of a Type Ia scenario, which I will briefly describe here.

The basic idea is that a dense wind has emanated  from the secondary star with  a mass-loss rate of $\dot{M} $~\msun\,yr$^{-1}$,
and a wind velocity $v_\mathrm{w}$. The wind must have lasted long enough for the wind material to reach the current
radius of Kepler's SNR ($2.6 d_5$~pc) implying that the wind  must have prolonged for
at least $t_\mathrm{w} \gtrsim 2.5 \times 10^5 d_5 (v_\mathrm{w}/10~\mathrm{km\,s}^{-1}$)~yr.
The total mass lost by the system  is given by 
\begin{equation}
M_\mathrm{wind} \gtrsim \dot{M}t_\mathrm{w}= \dot{M} \frac{R}{v_\mathrm{w}}=
2.5\left( 
\frac{\dot{M}}{10^{-5}\mathrm{M_\odot\,yr^{-1}}}\right)
\left(\frac{v_\mathrm{w}}{10\ {\rm km\,s}^{-1}}\right)^{-1}
d_5 \ \mathrm{M_\odot}.
\end{equation}
This means that the mass loss rate cannot have been much higher than $10^{-5}$~\msun\,yr$^{-1}$, as the total
mass loss of the donor could not exceed the mass of a viable donor star for a Type Ia progenitor,
which must lie in the range of 3-6~\msun.

The optical proper motions of the knots indicate a space of around 250-280~\kms\ (212 $d_5$~\kms in NW direction and 180 km/s
toward us). This likely corresponds to the space velocity of the progenitor system.  The combination of a spherical wind
and a space velocity gives rise to a bow shock around the system in the direction of the motion of the system.
The density of the wind as a function of radius $r$ is given by
\begin{equation}
\rho_\mathrm{w}(r)=\frac{\dot{M}}{4\pi r^2 v_\mathrm{w}} = 7.8\times 10^{-25}
\left(\frac{\dot{M}}{10^{-5}\,{\rm M_\odot}}\right) \left(\frac{v_\mathrm{w}}{10\ {\rm km\,s}^{-1}}\right)^{-1} \left(\frac{R}{2.6\,{\rm pc}}\right)^{-2}\, {\rm g\,cm}^{-3}.
\end{equation}
The densities in the Northwest appear to be  around 50~cm$^{-3}$ (\sect~\ref{sec:opt_ir}). The density in the wind
may be lower, as the gas encountered by the SNR shock may have been shocked by the wind termination shock,
and later by the SNR shock. However, we see that the mass loss rate cannot have been substantially lower
$10^{-5}$\msun\,yr$^{-1}$, as otherwise the wind densities would fall below the densities of the CSM component in Kepler's SNR.

The termination radius of the wind is now given by assuming  equilibrium between the ram
pressure of the wind $P_\mathrm{w}=\rho_\mathrm{w}v_\mathrm{w}^2$ and the ram
pressure of the interstellar medium, $P_\mathrm{ISM}=\rho_\mathrm{ISM}v_\mathrm{prog}^2$, with
$v_\mathrm{prog}$ the velocity of the progenitor system. Hence,
\begin{equation}
R_\mathrm{ts}=  1.9\ 
\left(\frac{\dot{M}}{10^{-5}\,{\rm M_\odot}}\right)^{1/2}
\left(
\frac{v_\mathrm{w}}{10\,\mathrm{km\,s^{-1}}}
\right)^{1/2} 
\left(\frac{v_\mathrm{prog}}{250\ \mathrm{km\,s^{-1}}}\right)^{-1}\left(\frac{n_\mathrm{ISM}}{10^{-3}\, \mathrm{cm^{-3}}}\right)^{-1/2}~\mathrm{pc},
\end{equation}
with $n_\mathrm{ISM}=10^{-3}$  an estimate for the ISM at 500~pc above the Galactic plane.
This estimate for the termination shock radius is in reasonable agreement with the radius of Kepler's SNR.

The simulations of \citet{chiotellis12} show that  both the bow-shock like shape of Kepler's SNR as well as the measured expansion parameters
can be reasonably well reproduced if SN\,1604 indeed went off inside a stellar wind bubble with a systematic velocity of 250~\kms. 
It was assumed that the ejecta mass was 1.4~\msun, typical for an exploding white dwarf. 
The simulations showed that for a distance of 4~kpc the radius of the termination shock must be around 2~pc. However,
a SNR with an \index{explosion energy} explosion energy of $10^{51}$~erg  going off inside such a smaller bubble, would have moved through the wind bubble
and the shock would have penetrated into the ISM, which is in disagreement with the observations. 
One can reproduce the characteristic of Kepler's SNR at this distance, but only if the explosion 
energy is reduced to $2\times 10^{50}$~erg. This is in disagreement with the typical  Type Ia explosion energies of $1.2 \times 10^{51}$~erg
\citep{woosley07}. Moreover, the high Fe content of Kepler's SNR rather suggest that SN\,1604 was a relatively energetic type Ia event
\citep{park13}. This problem does not occur if Kepler's SNR has a distance of $\gtrsim 6$~kpc.

A related hydrodynamical study by \citet{patnaude12}
concentrates  on the impact of hydrodynamics on the resulting X-ray spectra. The wind bubble evolution itself is not modelled, but  an 
$1/r^2$ density
distribution is assumed. They confirm that normal Type Ia explosion energies are not consistent with a distance of Kepler's SNR
of 4~kpc, but that instead a distance of  7~kpc should be considered. \citet{toledo14} follow more closely the scenario of \citet{chiotellis12},
but the simulations are in 3D and the wind loss is assumed to be anisotropic. As a result more structure is found in the simulated X-ray maps,
some of which may explain for example the central bar in Kepler's SNR (e.g. their Fig.~10). In general, the anisotropy of the wind
can result in a peculiar structure. \citet{burkey13} simulated the evolution of the SNR inside a wind strongly varying
as a function of the polar axis. However, the simulation does not take into account the effects of the velocity of the system as a whole,
and the expansion parameter of the plasma.

\section{The progenitor system of SN\,1604}

\subsection{Elevated circumstellar nitrogen abundances, silicates  and a single degenerate scenario for SN\,1604}
\citet{chiotellis12} argued that the mass loss properties inferred for the progenitor system of SN\,1604 suggest that the progenitor
system consisted of CO white dwarf accreting wind material from a Asymptotic Giant Branch (AGB) \index{Asymptotic Giant Branch (AGB) star} star. 
Indeed, the high mass
loss rates and the wind expansion time scale are in good agreement with what is generally inferred for AGB stars.
Given the relatively high total
mass loss ($\sim 2.5$\msun, plus matter that has accreted onto the white dwarf), 
it does mean that the donor star must have been a relatively massive star at the main sequence. 

In addition, the nitrogen-rich abundance of the wind,  and the silicate dust  (Sect.~\ref{sec:opt_ir}) also gives clues
about the donor star. The nitrogen-richness can be explained by \index{hot bottom burning} hot bottom burning \citep[e.g.][]{karakas10}, a process in which
hydrogen is burned through CNO cycle at the base of the outer convective envelope. This brings nitrogen to the surface.
Hot bottom burning occurs in stars with initial masses larger than 4-5~\msun.
Silicate dust \index{dust emission} \index{sillicates} predominantly forms in winds in which the carbon/oxygen ratio is smaller than 1. Otherwise the oxygen 
binds to carbon, making CO, leaving no oxygen for the build up of silicates. In this light it is interesting that \citet{mcsaveney07}
found two AGB stars in the Large Magellanic Clouds, which have elevated nitrogen abundance and a $C/O<1$. The masses
of these two stars are inferred to be 4~\msun\ and 6~\msun.

Taken all the evidence together we seem to have a rather consistent scenario for the progenitor system of SN\,1604 and the evolution
of the SNR: a single degenerate white dwarf, accreting wind material from a rather massive AGB star. The mass of the donor
then implies that the white dwarf was also rather massive at the main sequence (5-6~\msun). {Kepler's SNR provides therefore the
best case that at least some Type Ia supernovae are caused by single degenerate white dwarf systems.}

\subsection{Problems with a single degenerate Type Ia scenario for SN\,1604}

There two problems that need to be solved for the single degenerate scenario for SN\,1604, or even for any Type Ia binary scenario:

1) The donor star must be relatively massive (4-6\msun), and therefore a luminous star. A recent search by \citet{kerzendorf14} did not reveal
any bright enough star near the centre of Kepler's SNR. The former donor star \index{donor star} is likely to still be bright, although, its outer envelope
may have been removed by the supernova blast. Its space velocity should be close to that of the progenitor system.
The lack of a donor star argues instead for a double degenerate explosion for SN\,1604. A similar conclusion was
 drawn for other Type Ia SNRs, SNR 0509-675 and SN\,1006 \citep{schaefer12,gonzalez12}, \index{SNR G327.6+14.6}
for which also  no bright donor stars have been found. For Tycho's SNR a donor star has been identified \citep{ruiz04}, but this identification
is disputed \citep{kerzendorf13}. \index{SNR G120.1+1.4}

2) The velocity of the progenitor system must have been $\sim 250$~\kms. This implies that the system must have left
the Galactic plane around 3~million year ago. 
Such a time scale is too short for  any Type Ia scenario, which should have evolutionary time scales of $>50$ Myr \citep[e.g.][]{claeys14}.
 One can of course speculate that Kepler's SNR
is bound to the Galactic plane, and has been oscillating far up and below the Galactic plane for  considerable longer times.
This still creates, however, the problem of how to slingshot away a binary system with 250~\kms. Runaway stars \index{runaway stars}
are usually created by
close interactions of triple systems, but in 90\% of the cases single stars are being ejected,
and the highest velocities are not expected to be obtained by the ejected binaries \citep{leonhard90}. Moreover, in the AGB scenario
for SN\,1604 the binary system must have been relatively wide \citep{chiotellis12}. Nevertheless, observationally some high velocity binaries are known.
A well known example is the Mira system, which has a space velocity of about 100~\kms, and which consists also of an AGB star and
a probable white dwarf \citep{martin07}.

Problem 1) is a problem for only the single degenerate scenario, but problem 2) is a problem for any Type Ia  scenario involving
a binary system. 

\subsection{Was SN\,1604 a core-degenerate Type Ia explosion?}
\label{sec:core}
A possible solution to problem 1 is so-called core degenerate scenario \index{core degenerate scenario for Type Ia supernovae} 
\citep{ilkov12,tsebrenko13} according to which a single degenerate system
may form a common envelope at the end of the evolution of the secondary star (an AGB star). The white dwarf
and the core of the AGB star merge at the very end of the AGB star evolution, or shortly thereafter. The resulting white dwarf may
be super-Chandrasekhar, but gravitational collapse followed by an explosion can be delayed due to a rapid rotation of the massive white dwarf.
 Indeed, this scenario is somewhat
intermediate between the single degenerate and the double degenerate scenario, as the supernova is clearly caused by a double
degenerate merger, but the time scales involved can be  similar to the single degenerate scenario. A consequence of this scenario
is that the supernova mass does not have to be 1.38 \msun, but could be more. Another possible consequence for SN\,1604 may be that
there was high density material surrounding the progenitor, consisting of the envelope removed by the common envelope event. 
No light echo has yet been detected of SN\,1604 \citep{rest08}, but if it will be detected and an optical spectrum can be obtained,
perhaps we will learn more about the event itself and the presence of material in the immediate vicinity of the supernova.

\subsection{What can we learn from the historical light curve of SN\,1604?}
\label{sec:lightcurve}

Up to the 1990ies the hope was that the historical light curves \index{supernova light curve} \index{supernova light curve (SN 1604)} 
of SN\,1572 and SN\,1604 could shed  light on the maximum
brightness of Type Ia supernovae, and hence be used to constrain the Hubble parameter \citep[e.g.][]{danziger80,schaefer96}. 
Nowadays the Hubble parameter is known
to sufficient detail, but the historical light curves remain of interest, as it may lead to sub-typing of the historical supernovae and
provide estimates on the distance of  SNRs (\sect.~\ref{sec:properties}). \citet{baade43} was the first one to determine
the historical light curve of SN\,1604, and this was later augmented with magnitude estimates based on Korean observations
by \citet{clark77}. These data have been compared to supernovae light curve by among others \citet{schaefer96}, and most
recently \citet{katsuda15}.

\begin{figure}
\centerline{
\includegraphics[trim=50 50 50 50,clip=true,width=1.\textwidth]{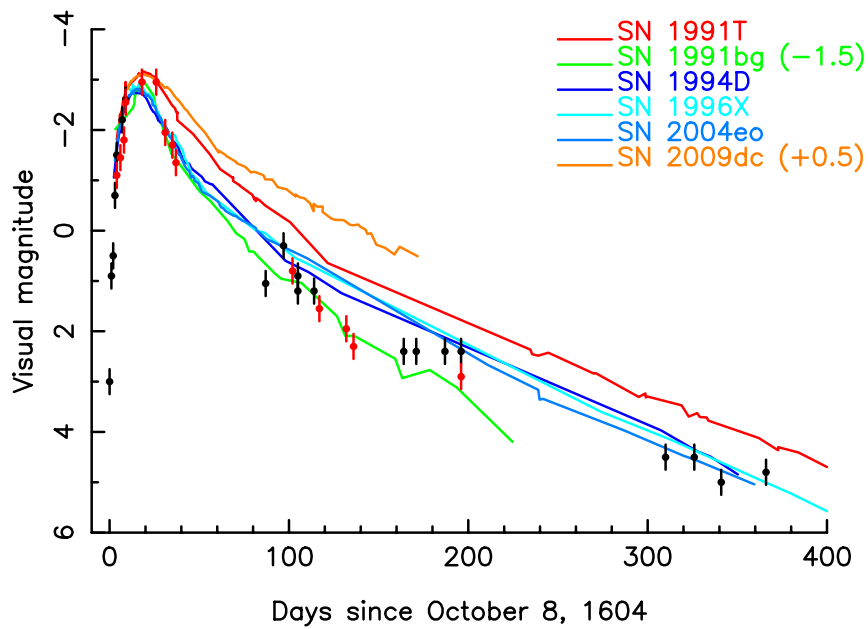}}
\caption{The light curve of SN\,1604 based on the European observations as interpreted by \citet[][black data points]{baade43} and
the Korean observations collected by \citet[][red data points]{stephenson02}.
For comparison the light curves of several other Type Ia supernovae have been over-plotted.
These  curves have been scaled to that of SN\,1604, assuming a distance for SN\,1604 of 5~kpc and $A_V=2.8$.
More or less normal Type Ia supernovae light curves have been assigned bluish colours.
See Table~\ref{tab:vmax} for details.
\label{fig:lc}
}
\end{figure}

The historical light curve of SN\,1604 is shown in Fig.~\ref{fig:lc} with a typical magnitude error of 0.25-0.5. The historical light curve
is compared to a number of recently observed Type Ia supernovae (Table~\ref{tab:vmax} for an overview). 
These lightcurves have been scaled to the light curve of SN\,1604
by taking into account the distance moduli ($\mu$) and visual extinction parameters ($A_V$) of the supernovae, and assuming
a distance of 5~kpc to SN\,1604 (somewhat in the middle of current estimates, Sect.~\ref{sec:properties}) and $A_V=2.8$ for the historical supernova.
The latter is based on the estimate of \citet{blair91} for the non-radiative shock \index{non-radiative shocks} 
emission. 

According to Type supernova light curve models, the peak magnitude of Type Ia supernovae is determined by
the amount of radioactive $^{56}$Ni produced in the explosion (typically $\sim 0.7$\msun), whereas the steepness
of the decline after maximum is determined by the diffusion of heat generated by radioactivity to the photosphere of
the supernova; the more mass the slower the decline \citep{arnett79,cappellaro97,dado15}. Hence,
the low peak magnitude and fast decline of SN\,1991bg has been attributed to a sub-Chandrasekhar Type Ia explosion \index{sub-Chandrasekhar Type Ia supernova}
\citep{turatto96,drout13}, whereas the very bright and slow declining SN\,2009dc is thought to indicate 
a super-Chandrasekhar \index{super-Chandrasekhar Type Ia supernova} Type Ia \citep{silverman11}. 
The latter has a light curve that is even broader
and brighter than SN\,1991T, and its light curve has to be scaled by $\Delta V=0.5$ in order to provide even an approximate
fit to SN\,1604.

Recently \citet{patnaude12} suggested that SN\,1604 may have been similar to the bright Type Ia SN\,1991T.
For a distance of 5~kpc the peak absolute peak magnitude of SN\,1991T and SN\,1604 match reasonably
well, but SN\,1991T has a much brighter late time light curve, and a slower post maximum decline. 
In order to make SN\,1604 as luminous as
SN\,1991T at late times,  the distance of SN\,1604 has to be $\sim$7~kpc.
So the light curve of SN\,1991T \index{SN 1991T} does not seem to be an overall good template for SN\,1604. 
The situation is even worse for the probable super-Chandrasekhar Type Ia SN,2009dc, which, 
even when scaled down in brightness,
shows a profile that is too broad to fit the historical light curve.

Normal Type Ia supernova light curves (SN\,1994D, SN\,1996X, SN\,2004eo) are in reasonable agreement with
the historical light curve of SN\,1604. In particular, they fit reasonably well around the peak and they agree
reasonably well at very late times (300--400 days), but they are brighter than SN\,1604 for days 100--200.

At the faint extreme, there are peculiar  Type Ia supernovae such as SN\,2009dc, whose light curve has to shifted up by
$\Delta V=1.5$ in Fig.~\ref{fig:lc} in order to have the peak brightness coincide with the peak of SN\,1604.
Such a shift would mean that SN\,1604 would be at a distance of 2.5~kpc,
which is lower than most independent distance estimates (Sect.~\ref{sec:properties}).
As \citet{katsuda15} showed, when properly scaled to match the peak magnitude of SN\,1604, 
the light curve of SN\,2009dc is rather similar to the light curve of SN\,1604,
as both are fastly declining in brightness after peak brightness. \index{SN 1991bg} \index{SN 2005ek}
A similarly fast decline Type Ia was SN\,2005ek, whose light curve was used to estimate the ejecta mass of 
0.3-0.7~\msun, and a kinetic energy of $(2-5)\times 10^{50}$~erg. However, SN\,2005ek was poor in $^{56}$Ni,
at odds with what we know of SN\,1604 (Sect.~\ref{sec:xrays}). 
Perhaps one can still fit SN\,1604 with a faint Type Ia model, if one allows for a small
(sub-Chandrasekhar) progenitor mass, but with a large fraction of $^{56}$Ni.
This would make a fainter supernova, with a rapid decline, but it could
still allow for the high iron fraction in Kepler's SNR.
The reason this may work is that in X-rays the fraction of iron can be much better
determined than the absolute mass of iron.
A consequence is then that the distance of Kepler's SNR is  less than 5~kpc,
as the supernova is fainter, but 
this can be  accommodated
with hydrodynamical models since the supernova is then probably
also a sub-energetic Type Ia \citep[$<10^{51}$~erg,][]{chiotellis12}.
 
So what can we conclude from the historical light curve, with some reserve as the magnitudes derived from the historical 
observations have to be treated with some caution, and the extinction is uncertain?
First of all, the light curves of normal Type Ia seem overall in better agreement with the light curve of SN\,1604 than
the light curve of SN\,1991T.
Secondly, the light curve of SN\,1604 may have declined rather rapidly after peak brightness. The evidence
for this is uncertain, but could indicate a sub-Chandrasekhar explosion. However,
in order to account for the high iron fraction of Kepler's SNR, SN\,1991bg and SN\,2005ek are probably
not the right models. Moreover, a sub-Chandrasekhar model, would result in a shorter distance of Kepler's SNR.

A possibility that needs to be explored better is
whether it is possible that the fast decline of SN\,1604 could perhaps be the result of an overall faster expansion, and/or
a distribution of nickel closer to the photosphere. This helps the heat generated by $^{56}$Ni to diffuse  faster to the photosphere,
resulting in a more rapid post-maximum decline.
In any case, the fast decline is inconsistent with a super-Chandrasekhar explosion. This is of interest given the suggestion that
SN\,1604 could provide a evidence for the core-degenerate model  \citep[Sect.~\ref{sec:core}][]{tsebrenko13},
which can result in super-Chandrasekhar Type Ia explosions.

\begin{table}
  
  \caption{A comparison of SN\,1604 to the Type Ia SNe shown in Fig.~\ref{fig:lc}. \label{tab:vmax}
  }
  \scriptsize
  \begin{tabular}{l rcrcll}\hline\hline\noalign{\smallskip}
    &	$V_\mathrm{max}$	&	$A_V$	&	$m-M$	&	$M_V$	&	Remark	&	References	\\
    \noalign{\smallskip}\hline\noalign{\smallskip}
    SN\,1604:&\\
    @ 3 kpc	&	-2.75	&	2.8	&	12.39	&	-17.94	&				&	{\scriptsize	\citet{blair91}	}	\\
&	-2.75	&	3.5	&	12.39	&	-18.64	&				&	{\scriptsize	\citet{danziger80}	}	\\
 @ 6 kpc	&	-2.75	&	2.8	&	13.89	&	-19.44	&				&	{\scriptsize	\citet{blair91}	}	\\
 	&	-2.75	&	3.5	&	13.89	&	-20.14	&				&	{\scriptsize	\citet{danziger80}	}	\\
\noalign{\smallskip}\hline\noalign{\smallskip}																	
SN\,1991T	&\ \ 11.50	& \ \ 0.37	& \ \ 	30.57	&\ \	-19.44 \ \ &	{\scriptsize	Bright Type Ia	}	&	{\scriptsize	\citet{sasdelli14}	}	\\
SN\,1991bg \index{SN 1991b}	&	13.96	& \ \ 	0.12	&	31.61	&	-17.77	&{\scriptsize	Sub-Chandrasekhar	}	&	{\scriptsize	\citet{turatto96,gomez04}	}	\\
SN\,1994D	&	11.90	& \ \ 	0.08	&	30.82	&	-19.00	&	{\scriptsize	Normal Type Ia	}	&	{\scriptsize	\citet{cappellaro97,feldmeier07}	}	\\
SN\,1996X	&	13.21	& \ \ 	0.22	&	31.15	&	-18.16	&	{\scriptsize	Normal Type Ia	}	&	{\scriptsize	\citet{salvo01}	}	\\
SN\,2004eo	&	15.35	&	0.34	&	34.12	&	-19.11	&	{\scriptsize	Sub-Chandrasekhar?	}	&	{\scriptsize	\citet{pastorello07}	}	\\
SN\,2009dc	&	15.35	& \ \ 	0.31	&	34.93	&	-19.89	&	{\scriptsize	Super-Chandrasekhar	}	&	{\scriptsize	\citet{silverman11}	}	\\
\noalign{\smallskip}\hline
  \end{tabular}
\end{table}

\section{Conclusions}

The remnant of SN\,1604, Kepler's SNR, is one of the most remarkable SNRs, with its high bulk velocity of ~250~\kms.
SN\,1604 is now generally regarded to  have been a Type Ia supernova, whose
remnant is interacting with wind material from its progenitor system. One can even infer that the donor star must have been a 4-6 \msun\ star that
had evolved to the AGB phase. But this single degenerate Type Ia scenario for SN\,1604 has only one problem: a surviving donor has not been detected.
A core-degenerate supernova scenario, may offer a viable alternative theory, but it has its own problems.
Whatever the explosion scenario, the high space velocity of the progenitor system remains mysterious, as it requires the ejection
of a binary system out of the Galactic plane with $\sim 250$~\kms.

In Johannes Kepler's age SN\,1604 was recognised as  a unique, mysterious event that was thought to have profound implications for mankind. 
We may no longer think that SN\,1604 is of prime importance for mankind, but SN\,1604 is of prime
importance for understanding Type Ia supernovae.  
SN\,1604 has, even more than 400 yr after discovery, still profound implications for understanding cosmic phenomena, and
it has still not shared all its mysteries with us.

\section*{acknowledgement}
This chapter is born out of a long fascination for Kepler's supernova remnant; and my view on the supernova remnant
was influenced by my collaboration with my former graduate students Alexandros Chiotellis and Klara Schure. 
During this collaboration also discussions with Onno Pols about the evolution of binary stars and the process of hot bottom
burning were much appreciated.

\printindex

\end{document}